\documentclass[a4paper,11pt]{article}
\usepackage{amsmath}
\usepackage{amssymb}

\begin{document}




\thispagestyle{empty}
\begin{titlepage}

\vspace*{-1cm}
\hfill \parbox{3.5cm}{BUTP-98/20 \\ hep-ph/9808387} \\
\vspace*{1.0cm}

\begin{center}
  {\large {\bf \hspace*{-0.2cm}Decay and scattering amplitudes are not the same}
      \footnote{Contribution to the Int. workshop "Hadronic atoms and positronium
      in the standard \hspace*{0.5cm} model" , Dubna, 26-31 May 1998.} '
      \footnote{Work
      supported in part by the Schweizerischer Nationalfonds.}  }
  \vspace*{3.0cm} \\

{\bf
     P. Minkowski} \\
     Institute for Theoretical Physics \\
     University of Bern \\
     CH - 3012 Bern , Switzerland
   \vspace*{0.8cm} \\  

3. August 1998

\vspace*{4.0cm}

\begin{abstract}
\noindent
We study the lowest lying $\pi^{+} \ \pi^{-}$
resonance R, bound by the long range Coulomb potential and destabilized
by short range strong interactions mediating the dominant decay into two neutral pions.
Parametrizing the corresponding partial decay width by

$\Gamma \ ( \ R \ \rightarrow \ 2 \pi^{0} \ ) \ =
\ \frac{2}{9} \ \alpha^{\ 3} \ \sqrt{ m_{\pi^{+}}^{2} - m_{\pi^{0}}^{2}}
\ \left ( \ \Delta \ a_{\ D} \ m_{\pi^{+}} \ \right )^{\ 2}$

\noindent
the quantity $\Delta \ a_{\ D}$ depends within QCD \& QED on four
basic parameters : $\Lambda_{QCD} \ , \ m_{u} + m_{d} \ , \ \alpha \ , \ m_{d} - m_{u}$.

\noindent
We are interested here only in the limit where the last two parameters vanish  

$\Delta \ a_{\ D}^{\ 0} \ = \
\ \Delta \ a_{\ D} \ ( \
\Lambda_{QCD} \ , \ m_{u} + m_{d} \ , \ \alpha = 0 \ , \ m_{d} - m_{u} = 0 \ )$.
\vspace*{0.1cm}

\noindent
While isospin invariance implies 
$\Delta \ a_{\ D}^{\ 0} \ = \ a_{\ D}^{\ I = 0} \ - \ a_{\ D}^{\ I = 2}$ 
it is shown that beyond the first order expansion in the strong interaction
the identities $a_{\ D}^{\ I} \ = \ a^{\ I}$, where $a^{\ I}$ denote the 
strong interaction scattering lengths, do not hold.
\end{abstract}
\end{center}

\end{titlepage}







\pagestyle{plain}




\noindent
\section {Introduction, general and specific issues}
\label{sec1}
\vspace*{0.1cm} 

\noindent
The general physics situation addressed here deals with the interplay
of a long range but weak binding force 
( specifically Coulomb attraction within $\pi^{-}$ mesonic atoms or pionium )
and strong but short range forces ( repulsive and/or attractive ) 
which however do not themselves give rise to binding
(specifically destabilising Coulombic bound states, which become resonances
  due to the strong transition $\pi^{-} X \ \rightarrow \ \pi^{0} \ Y$ in
  particular 
  $\pi^{-} \pi^{+} \ \rightarrow \ \pi^{0} \ \pi^{0}$).
\vspace*{0.1cm} 

\noindent
In this paper we will exclusively deal with the lowest lying $\pi^{-} \pi^{+}$
(pionium) resonance, denoted R, which in purely nonrelativistic Coulomb
spectroscopic terms corresponds to the 1S state of the composing charged pions.
Yet the above general situation not only applies also to $\pi^{-} p$ and
$\pi^{-} d$ systems, i.e. pionic hydrogen and deuterium \cite{pihyd} , \cite{pideut} ,
but applies as well to (almost) classical gravitationally bound systems,
e. g. Halley's comet, subject to the short distance solar radiation pressure,
which eventually will dissolve the bulk of the comet upon successive near approaches.
\vspace*{0.1cm} 

\noindent
For our system R, the (partial) width $\Gamma \ ( \ R \ \rightarrow \ 2 \pi^{0} \ )$
is given by the expression 

\begin{equation}
  \label{eq:1}
      \begin{array}{c} 
\Gamma \ ( \ R \ \rightarrow \ 2 \pi^{0} \ ) \ =
\ f_{B} 
\hspace*{0.2cm}
      \begin{array}{c} 
        q_{f}
        \vspace*{0.2cm} \\
        \hline \vspace*{-0.2cm} \\
        8 \ \pi \ m_{\ R}^{\ 2}
      \end{array}
\hspace*{0.2cm}
| \ T_{\ 2 \ \leftarrow \ R} \ |^{\ 2}
        \vspace*{0.5cm} \\
	q_{f} \ = 
\hspace*{0.2cm}
      \begin{array}{c} 
        \sqrt{\ m_{\ R}^{\ 2} \ -
	\ 4 \ m_{\ \pi_{\ 0}}^{\ 2} \ }
        \vspace*{0.2cm} \\
        \hline \vspace*{-0.2cm} \\
        2
      \end{array}
    \hspace*{0.3cm} ; \hspace*{0.3cm} 
    f_{B} \ = \ \frac{1}{2} 
\hspace*{0.2cm} : \hspace*{0.2cm} 
\mbox{Bose factor}
        \vspace*{0.5cm} \\
T_{\ 2 \ \leftarrow \ R} \ = 
\ \left \langle \ 2 \ \pi^{0} \ ; \ k_{\ 1} \ , \ k_{\ 2} \ \right |
\ T \ \left | \ R \ ; \ p \ \right \rangle
    \hspace*{0.3cm} ; \hspace*{0.3cm} 
    p \ = \ k_{\ 1} \ + \ k_{\ 2}
\end{array}
\end{equation}

\noindent
In eq. (\ref{eq:1}) 
state vectors are labelled by particle content in asymptotic outgoing plane waves,
characterized by four momentum vectors $k_{\ 1,2}$ and $p$ , subject to 
energy momentum conservation. The relativistic normalization of
one particle states is adopted throughout.
\vspace*{0.1cm} 

\noindent
$\left | \ R \ ; \ p \ \right \rangle$ denotes the pionium resonance state with total
energy momentum vector $p$ , which is unique modulo redundancies of relative
order \\
$\Gamma_{\ R} \ / \ m_{\ R} \ \sim \ 0.7 \ 10^{\ - 9}$.
\vspace*{0.1cm} 

\noindent
T denotes the momentum conservation reduced ${\bf T}$-matrix or transition operator

\begin{equation}
\begin{array}{l}
  \label{eq:2}
      \begin{array}{c} 
        {\bf S} \ - \ {\bf 1}
        \vspace*{0.2cm} \\
        \hline \vspace*{-0.2cm} \\
        i
      \end{array}
\hspace*{0.2cm}
= \ {\bf T} 
    \hspace*{0.3cm} ; \hspace*{0.1cm} 
\left \langle \ b \ ; \ p_{\ b} \ \right |
\ {\bf T} \ \left | \ a \ ; \ p_{ a} \ \right \rangle
\ = 
\begin{array}[t]{l}
( \ 2 \pi \ )^{\ 4} \ \delta^{\ 4} \ ( \ p_{\ b} \ - \ p_{ a} \ )
        \vspace*{0.5cm} \\
\ \left \langle \ b \ ; \ p_{\ b} \ \right |
\ T \ \left | \ a \ ; \ p_{ a} \ \right \rangle
\end{array}
\end{array}
\end{equation}

\noindent
It follows from the structure of the expression for the decay amplitude in 
eq. (\ref{eq:1}) , that the strong i.e. the short range interactions
dominate the transition matrix T on one hand, but also modify the
structure of the resonance. The latter is described by a convenient
wave function, best known from nonrelativistic interactions, represented
by potentials in all relevant channels. But a relativistic Bethe Salpeter
wave function or any variant thereof \cite{BS} exhibits the same modification
in principle. 
\vspace*{0.1cm} 

\noindent
The region, where the short range forces modify this wave function 
(in configuration space) is
small compared to the main Coulomb dominated volume proportional to the
cube of the pionium Bohr radius $ r_{\pi} \ = \ 2 \ / \ ( \ \alpha \ m_{\ \pi^{+}} \ )$.
But this reduced volume is {\em the dominant region} from where the decay
of the resonance takes place. 
\vspace*{0.1cm} 

\noindent
To lowest order in the strong interactions the resonance wave function remains
unmodified, but if higher orders are included, this modification becomes
important. Beyond this lowest order no obvious relation exists, even though higher order
isospin asymmetries and electromagnetic corrections are neglected,
which expresses the matrix element
in eq. (\ref{eq:1}) in terms of exclusively the strong scattering
amplitude of constituent pions on one hand and the purely Coulombic
wave function for the resonance on the other hand.
\vspace*{0.1cm} 

\noindent
Precisely such a relation has been derived by Deser, Goldberger, Baumann and Thirring
\cite{DGBT} for pionic atoms. Since its validity would encompass the much
more general interplay of long range weak binding and short range
strong but non binding forces, it can easily be falsified in potential
models. The fact, that such models may not be applicable
to pionium or more generally to pionic atoms is in this respect irrelevant.
\vspace*{0.1cm} 

\noindent
In section 2 we will derive such a relation involving beyond the
abovementioned quantities the dependence of the strong interaction scattering lengths
on an appropriate coupling parameter ( $\lambda$ ) for strong interactions

\begin{equation}
\begin{array}{l}
  \label{eq:3}
\Delta \ a_{\ D}^{\ 0} \ = \
\ \lambda 
\hspace*{0.2cm}
      \begin{array}{c} 
        d
        \vspace*{0.2cm} \\
        \hline \vspace*{-0.2cm} \\
        d \ \lambda 
      \end{array}
\hspace*{0.2cm}
\ \Delta \ a^{\ 0}
        \vspace*{0.5cm} \\
\ \Delta \ a^{\ 0} \ =
\ \left ( \ a^{\ I = 0} \ - \ a^{\ I = 2} \ \right )
\ ( 
\ \Lambda_{QCD}  ,  m_{u} + m_{d}  ,  \alpha = 0  ,  m_{d} - m_{u} = 0 \ )
\end{array}
\end{equation}

\noindent
In eq. (\ref{eq:3}) the quantities 
$\ \Delta \ a^{\ 0} $ , $ a^{\ I = 0,2} $ refer to the scattering length in
the limit specified.
\vspace*{0.1cm} 

\noindent
Estimates of the lifetime of pionium are presented in section 3.

\noindent
\section {Resonance decay amplitude in the Lippmann-Schwinger framework}
\label{sec2}
\vspace*{0.1cm} 

\noindent
The amplitude $T_{\ 2 \ \leftarrow \ R}$ in eq. (\ref{eq:1}) is
proportional to the reduced amplitude in relative space coordinates
as adapted to the center of mass frame 

\begin{equation}
\begin{array}{l}
  \label{eq:4}
T_{\ 2 \ \leftarrow \ R} \ = \ K
\ \left \langle \ \varphi_{\ \vec{k}}^{\ +} \ |
\ H_{\ 1} \ | \ R \ ; \ rel \ \right \rangle
    \hspace*{0.3cm} ; \hspace*{0.3cm} 
    \vec{k} \ = \ \frac{1}{2} \ ( \ \vec{k}_{\ 1} \ - \ \vec{k}_{\ 2} \ ) 
        \vspace*{0.5cm} \\
\left | \ \varphi_{\ \vec{k}}^{\ +} \ \right \rangle \ =
\ \left | \ 2 \ \pi^{0} \ , \ rel \ ; \ \vec{k} \ + \ \right \rangle
\end{array}
\end{equation}

\noindent
In eq. (\ref{eq:4}) $H_{\ 1}$ denotes the interaction Hamiltonian
in the Schr\"{o}dinger representation. The state
$ \left | \ R \ ; \ rel \ \right \rangle $ is normalized according to 
the scalar product in relative coordinate space. Because of the
open decay channel the resonance is an eigenstate in the continuous spectrum
of the total relative Hamiltonian, including all electromagnetic interactions 
\begin{equation}
\begin{array}{l}
  \label{eq:5}
 H_{\ rel} \ \left | \ R \ ; \ rel \ \right \rangle \ = 
\ m_{\ R} \ \left | \ R \ ; \ rel \ \right \rangle
    \hspace*{0.3cm} ; \hspace*{0.3cm} 
\ H_{\ rel} \ = \ H_{\ 0} \ + \ H_{\ 1} 
    \hspace*{0.3cm} \rightarrow \hspace*{0.3cm} 
    H
\end{array}
\end{equation}
\noindent
Furthermore $\varphi_{\ \vec{k}}^{\ +}$ in eq. (\ref{eq:4}) denotes the outgoing
scattering wave function obeying the Lippmann-Schwinger equation
\begin{equation}
\begin{array}{l}
  \label{eq:6}
\varphi_{\ \vec{k}}^{\ +}
\ =
\ \psi_{\ \vec{k}}
\ +
\ \begin{array}{c}
1
 \vspace*{0.3cm} \\
\hline	\vspace*{-0.3cm} \\
E \ - \ i \ \varepsilon \ - \ H
\end{array}
\hspace*{0.2cm} 
\ H_{\ 1} 
\ \psi_{\ \vec{k}}
        \vspace*{0.5cm} \\
	E \ = \ m_{\ R} \ = 2 \ \sqrt{\ k^{\ 2} \ + \ m_{\ \pi_{\ 0}}^{\ 2} \ }
    \hspace*{0.3cm} ; \hspace*{0.3cm} 
    \left \langle \ \vec{x} \ \right | \left .
\ \psi_{\ \vec{k}} \ \right \rangle \ = \ \exp \ ( \ i \ \vec{k} \ \vec{x} \ )
\end{array}
\end{equation}
\noindent
In eq. (\ref{eq:6}) $ \psi_{\ \vec{k}} $ denotes a plane wave.
\vspace*{0.1cm} 

\noindent
A fully relativistic
description of relative coordinate space involves the use of relative time
and the corresponding Bethe Salpeter equation \cite{BS} .
\vspace*{0.1cm} 

\noindent
Finally the kinematic constant K in eq. (\ref{eq:4}) is determined from 
the equivalent expression for the resonance width in eq. (\ref{eq:1})
\begin{equation}
  \label{eq:7}
      \begin{array}{c} 
\Gamma \ ( \ R \ \rightarrow \ 2 \pi^{0} \ ) \ =
 \begin{array}[t]{c}
\ f_{B} \ {\displaystyle{\int}} \ ( \ 2 \pi \ ) \ \delta \ ( \ E \ - \ m_{\ R} \ )
\hspace*{0.2cm} 
 \begin{array}{c}
d^{\ 3} \ k
 \vspace*{0.3cm} \\
\hline	\vspace*{-0.3cm} \\
( \ 2 \pi \ )^{\ 3}
\end{array}
\hspace*{0.2cm} 
        \vspace*{0.5cm} \\
\left | \ \left \langle \ \varphi_{\ \vec{k}}^{\ +} \ |
\ H_{\ 1} \ | \ R \ ; \ rel \ \right \rangle
\ \right |^{\ 2}
\end{array}
\end{array}
\end{equation}
\noindent
From eq. (\ref{eq:7}) we determine the constant K in eq. (\ref{eq:4})
\begin{equation}
  \label{eq:8}
      \begin{array}{c} 
\Gamma \ ( \ R \ \rightarrow \ 2 \pi^{0} \ ) \ =
\ f_{B} 
\hspace*{0.2cm}
      \begin{array}{c} 
        q_{f} \ m_{\ R}
        \vspace*{0.2cm} \\
        \hline \vspace*{-0.2cm} \\
        4 \ \pi 
      \end{array}
\hspace*{0.2cm}
\ \left | \ \left \langle \ \varphi_{\ \vec{k}}^{\ +} \ |
\ H_{\ 1} \ | \ R \ ; \ rel \ \right \rangle
\ \right |^{\ 2}
        \vspace*{0.5cm} \\
	K \ = \ \sqrt{2} \ ( \ m_{\ R} \ )^{\ 3/2}
\end{array}
\end{equation}
\vspace*{0.1cm} 

\noindent
{\bf Associated $2 \ ' \ \rightarrow \ 2$ scattering amplitudes}
\vspace*{0.1cm} 

\noindent
The resonance decay amplitude in eqs. (\ref{eq:1}) and (\ref{eq:4})
is associated with the scattering amplitude for the $2' \ \rightarrow \ 2$
pion reaction
\begin{equation}
  \label{eq:9}
      \begin{array}{c} 
      \begin{array}{lll ll} 
      \pi^{0}  & \pi^{0} &            & \pi^{+} & \pi^{-} \\
	       &         & \leftarrow &         &         \\
       k_{\ 1} & k_{\ 2} &            & p_{\ 1} & p_{\ 2} 
\end{array}
        \vspace*{0.5cm} \\
      \begin{array}{c} 
       d \ \sigma \ ( \ 2 \ ' \ \rightarrow \ 2 \ ) 
        \vspace*{0.2cm} \\
        \hline \vspace*{-0.2cm} \\
       d \ \Omega 
      \end{array}
\hspace*{0.2cm}
= 
\ f_{B} 
\ \left (
\hspace*{0.2cm}
      \begin{array}{c} 
        q_{f}
        \vspace*{0.2cm} \\
        \hline \vspace*{-0.2cm} \\
        q_{i} 
      \end{array}
\hspace*{0.2cm}
\right )_{\ cm}
\hspace*{0.2cm}
      \begin{array}{c} 
        1
        \vspace*{0.2cm} \\
        \hline \vspace*{-0.2cm} \\
        64 \ \pi^{\ 2} \ s
      \end{array}
\hspace*{0.2cm}
| \ T_{\ 2 \ \leftarrow \ 2 \ '} \ |^{\ 2}
        \vspace*{0.5cm} \\
T_{\ 2 \ \leftarrow \ 2 \ '}
\ =
\ \left \langle \ 2 \ \pi^{0} \ ; \ k_{\ 1} \ , \ k_{\ 2} \ |
\ T \ | \ \pi^{+} \ \pi^{-} \ ; \ p_{\ 1} \ , \ p_{\ 2} \ \right \rangle
\end{array}
\end{equation}
\noindent
All kinemaric quantities in eq. (\ref{eq:9}) refer to the center of mass system.

\noindent
The invariant amplitude $ T_{\ 2 \ \leftarrow \ 2 \ '} $ depends on the
standard Lorentz invariants 
\begin{equation}
  \label{eq:10}
      \begin{array}{c} 
s \ = \ ( \ p_{\ 1} \ + \ p_{\ 2} \ )^{\ 2}
    \hspace*{0.3cm} ; \hspace*{0.3cm}
t \ = \ ( \ p_{\ 1} \ - \ k_{\ 1} \ )^{\ 2}
    \hspace*{0.3cm} ; \hspace*{0.3cm}
u \ = \ ( \ p_{\ 1} \ - \ k_{\ 2} \ )^{\ 2}
        \vspace*{0.5cm} \\
	k_{\ 1} \ + \ k_{\ 2} \ = \ p_{\ 1} \ + \ p_{\ 2} \ =
\ ( \ E \ , \ \vec{0} \ )
        \vspace*{0.5cm} \\
	q_{f} \ = 
\hspace*{0.2cm}
      \begin{array}{c} 
        \sqrt{\ s \ -
	\ 4 \ m_{\ \pi_{\ 0}}^{\ 2} \ }
        \vspace*{0.2cm} \\
        \hline \vspace*{-0.2cm} \\
        2
      \end{array}
    \hspace*{0.3cm} ; \hspace*{0.3cm}
	q_{i} \ = 
\hspace*{0.2cm}
      \begin{array}{c} 
        \sqrt{\ s \ -
	\ 4 \ m_{\ \pi_{\ +}}^{\ 2} \ }
        \vspace*{0.2cm} \\
        \hline \vspace*{-0.2cm} \\
        2
      \end{array}
\end{array}
\end{equation}

\noindent
Analogous to the relative coordinate space decay amplitude \\
$\left \langle \ \varphi_{\ \vec{k}}^{\ +} 
\ | \ H_{\ 1} \ | \ R \ ; \ rel \ \right \rangle$
in eq. (\ref{eq:4}) the corresponding $2 \ ' \ \rightarrow \ 2$ 
scattering amplitude is

\begin{equation}
  \label{eq:11}
\begin{array}{l}
T_{\ 2 \ \leftarrow \ 2 \ ' } \ = \ K \ '
\ \left \langle \ \varphi_{\ \vec{k}}^{\ +} \ |
\ H_{\ 1} \ | \ \psi_{\ \vec{p}} \ \right \rangle
    \hspace*{0.3cm} ; \hspace*{0.3cm} 
    \vec{k} \ = \ \frac{1}{2} \ ( \ \vec{k}_{\ 1} \ - \ \vec{k}_{\ 2} \ ) 
        \vspace*{0.5cm} \\
\left | \ \varphi_{\ \vec{k}}^{\ +} \ \right \rangle \ =
\ \left | \ 2 \ \pi^{0} \ , \ rel \ ; \ \vec{k} \ + \ \right \rangle
        \vspace*{0.5cm} \\
    \vec{p} \ = \ \frac{1}{2} \ ( \ \vec{p}_{\ 1} \ - \ \vec{p}_{\ 2} \ ) 
    \hspace*{0.3cm} ; \hspace*{0.3cm} 
\left | \ \psi_{\ \vec{p}} \ \right \rangle 
\ =
\ \left | \ \pi^{+} \ \pi^{-} \ , \ rel \ ; \ \vec{p} \ \right \rangle
\end{array}
\end{equation}

\noindent
$\left | \ \psi_{\ \vec{p}} \ \right \rangle$ in eq. (\ref{eq:11})
refers to the plane wave asymptotic $\pi^{+} \ \pi^{-}$ state.
\vspace*{0.1cm} 

\noindent
The expression for the cross section in eq. (\ref{eq:9}) becomes
\begin{equation}
  \label{eq:12}
\begin{array}{l}
d \ \sigma \ ( \ 2 \ ' \ \rightarrow \ 2 \ ) 
\ = 
\begin{array}[t]{c}
\ f_{B} \ {\displaystyle{\int}} \ ( \ 2 \pi \ ) \ \delta \ ( \ E \ - \ E \ ' \ )
\hspace*{0.2cm} 
 \begin{array}{c}
d^{\ 3} \ k
 \vspace*{0.3cm} \\
\hline	\vspace*{-0.3cm} \\
( \ 2 \pi \ )^{\ 3}
\end{array}
\hspace*{0.2cm} 
        \vspace*{0.5cm} \\
\ \left |
\ \left \langle \ \varphi_{\ \vec{k}}^{\ +} \ |
\ H_{\ 1} \ | \ \psi_{\ \vec{p}} \ \right \rangle
\ \right |^{\ 2}
\hspace*{0.2cm} 
 \begin{array}{c}
1
 \vspace*{0.3cm} \\
\hline	\vspace*{-0.3cm} \\
v_{\ lab}
\end{array}
\hspace*{0.2cm} 
\end{array}
        \vspace*{0.5cm} \\
	E \ = \ 2 \ \sqrt{\ \vspace*{0.1cm} 
 \begin{array}{c}
	k^{\ 2} \ + \ m_{\ \pi_{\ 0}}^{\ 2} 
\end{array}
	 \ }
    \hspace*{0.3cm} ; \hspace*{0.3cm} 
	E \ '  \ = \ 2 \ \sqrt{\ p^{\ 2} \ + \ m_{\ \pi_{\ +}}^{\ 2} \ }
        \vspace*{0.5cm} \\
    v_{\ lab} \ =  
\hspace*{0.2cm} 
 \begin{array}{c}
2 \ p \ E
 \vspace*{0.3cm} \\
\hline	\vspace*{-0.3cm} \\
E^{\ 2} \ - \ 2 \ m_{\ \pi_{\ +}}^{\ 2}
\end{array}
\hspace*{0.2cm} 
\end{array}
\end{equation}

\noindent
Thus we obtain the expression for the differential cross section
equivalent to eq. (\ref{eq:9})
\begin{equation}
  \label{eq:13}
\begin{array}{l}
      \begin{array}{c} 
       d \ \sigma \ ( \ 2 \ ' \ \rightarrow \ 2 \ ) 
        \vspace*{0.2cm} \\
        \hline \vspace*{-0.2cm} \\
       d \ \Omega 
      \end{array}
\hspace*{0.2cm}
= 
\ \begin{array}[t]{c}
 f_{B} 
\ \left (
\hspace*{0.2cm}
      \begin{array}{c} 
        q_{f}
        \vspace*{0.2cm} \\
        \hline \vspace*{-0.2cm} \\
        q_{i} 
      \end{array}
\hspace*{0.2cm}
\right )_{\ cm}
\hspace*{0.2cm}
      \begin{array}{c} 
        1
        \vspace*{0.2cm} \\
        \hline \vspace*{-0.2cm} \\
        16 \ \pi^{\ 2} 
      \end{array}
\hspace*{0.2cm}
        \vspace*{0.5cm} \\
      \begin{array}{c} 
        E^{\ 2} \ - \ 2 \ m_{\ \pi_{\ +}}^{\ 2}
        \vspace*{0.2cm} \\
        \hline \vspace*{-0.2cm} \\
        2
      \end{array}
\hspace*{0.2cm}
\ \left |
\ \left \langle \ \varphi_{\ \vec{k}}^{\ +} \ |
\ H_{\ 1} \ | \ \psi_{\ \vec{p}} \ \right \rangle
\ \right |^{\ 2}
      \end{array}
\end{array}
\end{equation}

\noindent
Parametrizing the diferential cross section in the form 
\begin{equation}
  \label{eq:14}
\begin{array}{l}
      \begin{array}{c} 
       d \ \sigma \ ( \ 2 \ ' \ \rightarrow \ 2 \ ) 
        \vspace*{0.2cm} \\
        \hline \vspace*{-0.2cm} \\
       d \ \Omega 
      \end{array}
\hspace*{0.2cm}
= 
\ f_{B} 
\ \left (
\hspace*{0.2cm}
      \begin{array}{c} 
        q_{f}
        \vspace*{0.2cm} \\
        \hline \vspace*{-0.2cm} \\
        q_{i} 
      \end{array}
\hspace*{0.2cm}
\right )_{\ cm}
\ \left | \ F \ \right |^{\ 2}
\end{array}
\end{equation}
\noindent
and comparing eqs. (\ref{eq:9}) and (\ref{eq:14}) it follows
\begin{equation}
  \label{eq:15}
\begin{array}{l}
\begin{array}{lll}
F \ & = &
\hspace*{0.2cm}
      \begin{array}{c} 
        1
        \vspace*{0.2cm} \\
        \hline \vspace*{-0.2cm} \\
        8 \ \pi \ E 
      \end{array}
\hspace*{0.2cm}
\ T_{\ 2 \ \leftarrow \ 2 \ '}
        \vspace*{0.5cm} \\
\ & = &
\hspace*{0.2cm}
      \begin{array}{c} 
        m_{\ \pi_{\ +}}
        \vspace*{0.2cm} \\
        \hline \vspace*{-0.2cm} \\
        4 \ \pi  
      \end{array}
\hspace*{0.2cm}
\sqrt{\ 1 \ + 
     \ \begin{array}{c} 
 2 \ p^{\ 2}
        \vspace*{0.2cm} \\
        \hline \vspace*{-0.2cm} \\
  m_{\ \pi_{\ +}}^{\ 2} 
      \end{array}
 \ }
\ \left \langle \ \varphi_{\ \vec{k}}^{\ +} \ |
\ H_{\ 1} \ | \ \psi_{\ \vec{p}} \ \right \rangle
\end{array}
\end{array}
\end{equation}
\noindent
From eq. (\ref{eq:15}) we deduce the kinematical constant K '
\begin{equation}
  \label{eq:16}
\begin{array}{l}
K \ ' \ = \ 2 \ m_{\ \pi_{\ +}} \ E
\hspace*{0.2cm}
\sqrt{\ 1 \ + 
     \ \begin{array}{c} 
 2 \ p^{\ 2}
        \vspace*{0.2cm} \\
        \hline \vspace*{-0.2cm} \\
  m_{\ \pi_{\ +}}^{\ 2} 
      \end{array}
 \ }
    \hspace*{0.3cm} ; \hspace*{0.3cm}
    E \ = \ \sqrt{s}
\end{array}
\end{equation}
\noindent
The nonrelativistic limit involves the shift of the energy 
$E \ \rightarrow \ E \ - \ 2 \ m_{\ \pi_{\ +}}$
to zero value at $\pi^{+} \ \pi^{-}$ threshold.
\vspace*{0.1cm} 

\noindent
\section {Relations between decay and scattering amplitudes}
\label{sec3}
\vspace*{0.1cm} 

\noindent
We extrapolate the rate formula from above threshold to the resonance 
position with an arbitrary incident intensity I
\begin{equation}
  \label{eq:17}
\begin{array}{l}
I \ = \ v_{\ lab} \ \varrho 
    \hspace*{0.3cm} ; \hspace*{0.3cm}
\varrho \ = \left | \ \chi \ \right |^{\ 2} \ ( \ E \ )
        \vspace*{0.5cm} \\
\begin{array}{lll}
\Gamma \ ( \ E \ ; \ \varrho \ \rightarrow \ 2 \pi^{0} \ ) \ & = &
\ {\displaystyle{\int}} \ d \ \Omega
\hspace*{0.2cm}
      \begin{array}{c} 
       d \ \sigma \ ( \ 2 \ ' \ \rightarrow \ 2 \ ) 
        \vspace*{0.2cm} \\
        \hline \vspace*{-0.2cm} \\
       d \ \Omega 
      \end{array}
\hspace*{0.2cm}
I
        \vspace*{0.5cm} \\
	& = &
\ f_{B} 
\hspace*{0.2cm}
      \begin{array}{c} 
        q_{f} \ E
        \vspace*{0.2cm} \\
        \hline \vspace*{-0.2cm} \\
        4 \ \pi 
      \end{array}
\hspace*{0.2cm}
\hspace*{0.2cm}
\ \left |
\ \left \langle \ \varphi_{\ \vec{k}}^{\ +} \ |
\ H_{\ 1} \ | \ \psi_{\ \vec{p}} \ \right \rangle \ \chi
\ \right |^{\ 2}
\end{array}
\end{array}
\end{equation}
\noindent
to $E \ \rightarrow \ m_{\ R}$ , i.e. below $\pi^{+} \ \pi^{-}$ threshold.
It follows, that the resonance is described equivalently by the
appropriate choice of the amplitude $\chi \ \rightarrow \ \chi_{\ R}$
from eq. (\ref{eq:8})
\begin{equation}
  \label{eq:18}
\begin{array}{l}
\left \langle \ \varphi_{\ \vec{k}}^{\ +} \ |
\ H_{\ 1} \ | \ R \ ; \ rel \ \right \rangle
\ = 
\ \chi_{\ R}
\ \lim_{\ E \ \rightarrow \ m_{\ R}}
\ \left \langle \ \varphi_{\ \vec{k}}^{\ +} \ |
\ H_{\ 1} \ | \ \psi_{\ \vec{p}} \ \right \rangle
        \vspace*{0.5cm} \\
	p^{\ 2} \ \rightarrow
\hspace*{0.2cm}
      \begin{array}{c} 
        m_{R}^{\ 2} \ - \ 4 \ m_{\ \pi_{\ +}}^{\ 2}
        \vspace*{0.2cm} \\
        \hline \vspace*{-0.2cm} \\
        4 
      \end{array}
\hspace*{0.2cm}
\ < \ 0
\end{array}
\end{equation}
\noindent
We define the extrapolated scattering length $a_{\ R}$ below
$\pi^{+} \ \pi^{-}$ threshold in accordance with eq. (\ref{eq:15})
\begin{equation}
  \label{eq:19}
\begin{array}{l}
\ \lim_{\ E \ \rightarrow \ m_{\ R}}
\ \left \langle \ \varphi_{\ \vec{k}}^{\ +} \ |
\ H_{\ 1} \ | \ \psi_{\ \vec{p}} \ \right \rangle
\ =
\hspace*{0.2cm}
      \begin{array}{c} 
        4 \ \pi 
        \vspace*{0.2cm} \\
        \hline \vspace*{-0.2cm} \\
        m_{\ \pi_{\ +}} 
      \end{array}
\hspace*{0.2cm}
a_{\ R} \ \sqrt{2}
        \vspace*{0.5cm} \\
a_{\ R} \ = \ a_{\ R} 
\ \left ( \ " \ 2 \ \pi^{\ 0} \ " \ \leftarrow \ \pi^{\ +} \pi^{\ -} \ \right )
\end{array}
\end{equation}
\noindent
The combinatorial factor $\sqrt{2}$ included in the definition of $a_{\ R}$
in eq. (\ref{eq:19}) accounts for the two identical pions in an algebraic way.
The quotes in $ " \ 2 \ \pi^{\ 0} \ "$ refer to the bose symmetrized and normalized
state.
\vspace*{0.1cm} 

\noindent
Substituting $a_{\ R}$ in eq. (\ref{eq:17}) we obtain upon extrapolating
to the mass of the resonance 
\begin{equation}
  \label{eq:20}
      \begin{array}{c} 
\Gamma \ ( \ R \ \rightarrow \ 2 \pi^{0} \ ) \ =
\hspace*{0.2cm}
      \begin{array}{c} 
4 \ \pi \ q_{f} \ m_{\ R} 
        \vspace*{0.2cm} \\
        \hline \vspace*{-0.2cm} \\
        m_{\ \pi_{\ +}}^{\ 2} 
      \end{array}
\hspace*{0.2cm}
\ \left | 
\ a_{\ R} \ \chi_{\ R}
\ \right |^{\ 2}
\end{array}
\end{equation}
\noindent
It is rewarding in view of sequential approximations to use as unit of
density the Coulomb density $| \ \chi_{\ C} \ |^{\ 2}$ for the $\pi^{+} \ \pi^{-}$
system
\begin{equation}
  \label{eq:21}
      \begin{array}{c} 
\chi_{\ C} \ =
\ \left (
\ \alpha^{\ 3}
\hspace*{0.2cm}
      \begin{array}{c} 
        m_{\ \pi_{\ +}}^{\ 3} 
        \vspace*{0.2cm} \\
        \hline \vspace*{-0.2cm} \\
        8 \ \pi
      \end{array}
\hspace*{0.2cm}
\ \right )^{\ 1/2}
    \hspace*{0.3cm} ; \hspace*{0.3cm}
    \xi_{\ R} \ =
\hspace*{0.2cm}
      \begin{array}{c} 
        \chi_{\ R} 
        \vspace*{0.2cm} \\
        \hline \vspace*{-0.2cm} \\
        \chi_{\ C} 
      \end{array}
\hspace*{0.2cm}
\end{array}
\end{equation}
\noindent
The expression for the resonance width in eq. (\ref{eq:20}) becomes
\begin{equation}
  \label{eq:22}
      \begin{array}{c} 
\Gamma \ ( \ R \ \rightarrow \ 2 \pi^{0} \ ) \ =
\ \alpha^{\ 3}
\ q_{f}
\hspace*{0.2cm}
      \begin{array}{c} 
        m_{\ R} 
        \vspace*{0.2cm} \\
        \hline \vspace*{-0.2cm} \\
        2 \ m_{\ \pi_{\ +}} 
      \end{array}
\hspace*{0.2cm}
\ \left | 
\ m_{\ \pi_{\ +}} 
\ a_{\ R} \ \xi_{\ R}
\ \right |^{\ 2}
        \vspace*{0.5cm} \\
	a_{\ D} \ = \ a_{\ R} \ \xi_{\ R}
\end{array}
\end{equation}
\noindent
In eq. (\ref{eq:22}) the quantity $a_{\ D}$ denotes the decay equivalent
scattering length, to be distinguished from $a_{\ R}$ which is indeed
the scattering length, extrapolated to the resonance mass and for the
reaction $" \ 2 \ \pi^{0} \ \leftarrow \ \pi^{+} \ \pi^{-}$.
\vspace*{0.1cm} 

\noindent
Up to this point all equations were exact, to be evaluated in 
QCD , QED theory, restricted to two light quark flavors with 
masses $m_{\ u} \ , \ m_{\ d}$. 

\noindent
Hence $a_{\ D}$ is a function of 4 basic parameters 
\begin{equation}
  \label{eq:23}
      \begin{array}{c} 
	a_{\ D} \ =
	\ a_{\ D} \ \left (
	\ \Lambda_{\ QCD} \ ,
	\ m_{\ u} \ + \ m_{\ d} \ ,
	\ \alpha \ ,
	\ m_{\ d} \ - \ m_{\ u}
	\ \right )
\end{array}
\end{equation}
\noindent
We are only interested here in the limit, where the last two of the four
parameters in eq. (\ref{eq:23}) $\alpha$ and $m_{\ d} \ - \ m_{\ u}$
tend to zero, whereby only the lowest order contributions to
the resonance width $\Gamma \ ( \ R \ \rightarrow \ 2 \pi^{0} \ )$
are retained :
\begin{equation}
  \label{eq:24}
      \begin{array}{l} 
      \alpha \ , \ m_{\ d} - \ m_{\ u} \ \rightarrow \ 0
    \hspace*{0.3cm} ; \hspace*{0.3cm}
    m_{\ \pi_{\ 0}} \ \rightarrow \ m_{\ \pi_{\ +}} 
        \vspace*{0.5cm} \\
\Gamma \ ( \ R \ \rightarrow \ 2 \pi^{0} \ ) \ \rightarrow 
\Gamma^{\ 0} \ ( \ R \ \rightarrow \ 2 \pi^{0} \ )
    \hspace*{0.3cm} ; \hspace*{0.3cm}
    m_{\ R} \ \rightarrow \ 2 \ m_{\ \pi_{\ +}} 
        \vspace*{0.5cm} \\
	a_{\ D} \ \rightarrow
	\ a_{\ D}^{\ 0} \ =
	\ a_{\ D} \ \left (
	 \Lambda_{\ QCD} ,
	 m_{\ u} \ + \ m_{\ d} ,
	\ \alpha \ = \ 0 \ ,
	\ m_{\ d} \ - \ m_{\ u} \ = \ 0
	\ \right )
\end{array}
\end{equation}
\noindent
The momentum variables $\vec{p} \ , \ \vec{k}$ also tend to zero in the above limit
according to eq. (\ref{eq:18}).
\vspace*{0.1cm} 

\noindent
Before going over to the systematic approximation deffined in eq. (\ref{eq:24})
we represent the resonance width
$\Gamma \ ( \ R \ \rightarrow \ 2 \pi^{0} \ )$ in the form
\begin{equation}
  \label{eq:25}
      \begin{array}{l} 
\Gamma \ ( \ R \ \rightarrow \ 2 \pi^{0} \ ) \ =
\ \Gamma^{\ 0} \ ( \ R \ \rightarrow \ 2 \pi^{0} \ )
\ ( \ 1 \ + \ \delta \ )
        \vspace*{0.5cm} \\
	\delta \ = \ \delta \ ( \ \alpha \ , \ m_{\ d} \ - \ m_{\ u} \ )
	\ =
	\ O \ ( \ \alpha \ , \ ( \ m_{\ d} - \ m_{\ u} \ )^{\ 2} \ )
\end{array}
\end{equation}
\noindent
In eq. (\ref{eq:25}) only the relevant expansion parameters of the
correction factor $\delta$ are exhibited explicitly.
\vspace*{0.1cm} 

\noindent
The dependence of $\delta$ dominated by the first order $\alpha$ correction
has been discussed in this workshop by A. Rusetsky and H. Sazdjan
\cite{AR} , \cite{HSaz} and amounts numerically to 
\begin{equation}
  \label{eq:26}
      \begin{array}{l} 
      \delta \ \sim \ 0.06
\end{array}
\end{equation}
\noindent
enhancing in lowest nontrivial order the limiting width 
$ \Gamma^{\ 0} \ ( \ R \ \rightarrow \ 2 \pi^{0} \ )$ , to which
exclusive attention is directed in the following.
\vspace*{0.1cm} 

\noindent
{\bf The limiting situation : $\alpha \ , \ m_{\ d} - \ m_{\ u} \ \rightarrow \ 0$}
\vspace*{0.1cm} 

\noindent
From eqs. (\ref{eq:22}) and (\ref{eq:24}) we deduce the following
expression for the limiting width
$ \Gamma^{\ 0} \ ( \ R \ \rightarrow \ 2 \pi^{0} \ )$ 
\begin{equation}
  \label{eq:27}
      \begin{array}{l} 
\Gamma^{\ 0} \ ( \ R \ \rightarrow \ 2 \pi^{0} \ )
\ =
\ \alpha^{\ 3}
\ \sqrt{\ m_{\ \pi_{+}}^{2} - m_{\ \pi_{0}}^{2} \ }
\ \left | 
\ m_{\ \pi_{\ +}} 
	\ a_{\ D}^{\ 0}
\ \right |^{\ 2}
        \vspace*{0.5cm} \\
	a_{\ D}^{\ 0} \ = \ a_{\ R}^{\ 0} \ \xi_{\ R}^{\ 0}
\end{array}
\end{equation}
\noindent
In the limiting situation u-d isospin is an exact symmetry, which
implies the general decomposition into $I \ = \ 0 \ , \ 2$ channels
remembering the channel definition in eq. (\ref{eq:19})
\begin{equation}
  \label{eq:28}
      \begin{array}{l} 
a_{\ D \ , \ R}^{\ 0} \ = \ a_{\ D \ , \ R}^{\ 0} 
\ \left ( \ " \ 2 \ \pi^{\ 0} \ " \ \leftarrow \ \pi^{\ +} \pi^{\ -} \ \right )
        \vspace*{0.5cm} \\
      \begin{array}{lll} 
 \left | \ \pi^{\ +} \pi^{\ -} \ \right \rangle 
\ & = &
\ \sqrt{\ \frac{2}{3} \ }
\ \left | \ I \ = \ 0 \ \right \rangle 
\ -
\ \sqrt{\ \frac{1}{3} \ }
\ \left | \ I \ = \ 2 \ \right \rangle 
        \vspace*{0.5cm} \\
\left | \  " \ 2 \ \pi^{\ 0} \ " \ \right \rangle
\ & = &
\ \sqrt{\ \frac{1}{3} \ }
\ \left | \ I \ = \ 0 \ \right \rangle 
\ +
\ \sqrt{\ \frac{2}{3} \ }
\ \left | \ I \ = \ 2 \ \right \rangle 
\end{array}
\end{array}
\end{equation}
\noindent
It follows from eq. (\ref{eq:28}) relinquishing the subscript R on
$a_{\ R}^{\ 0}$ , the scattering length 
\begin{equation}
  \label{eq:29}
      \begin{array}{l} 
a_{\ R}^{\ 0} \ \rightarrow \ a^{\ 0} 
    \hspace*{0.3cm} : \hspace*{0.3cm}
a^{\ 0} \ = 
\hspace*{0.2cm}
      \begin{array}{c} 
        \sqrt{2}
        \vspace*{0.2cm} \\
        \hline \vspace*{-0.2cm} \\
        3
      \end{array}
\hspace*{0.2cm}
\Delta \ a^{\ 0}
    \hspace*{0.3cm} ; \hspace*{0.3cm}
a_{\ D}^{\ 0} \ = 
\hspace*{0.2cm}
      \begin{array}{c} 
        \sqrt{2}
        \vspace*{0.2cm} \\
        \hline \vspace*{-0.2cm} \\
        3
      \end{array}
\hspace*{0.2cm}
\Delta \ a_{\ D}^{\ 0}
        \vspace*{0.5cm} \\
\Delta \ a^{\ 0} \ = \ a^{\ I = 0} \ - \ a^{\ I = 2} 
    \hspace*{0.3cm} ; \hspace*{0.3cm}
\Delta \ a_{\ D}^{\ 0} \ = \ a_{\ D}^{\ I = 0} \ - \ a_{\ D}^{\ I = 2} 
\end{array}
\end{equation}
\noindent
In the following we drop also the superscript $^{0}$ on the quantities \\
$\Delta \ a^{\ 0} \ \rightarrow \ \Delta \ a$ and 
$\Delta \ a_{\ D}^{\ 0} \ \rightarrow \Delta \ a_{\ D}$ .
\vspace*{0.1cm} 

\noindent
Substituting eq. (\ref{eq:29}) in eq. (\ref{eq:27}) we obtain
\begin{equation}
  \label{eq:30}
      \begin{array}{l} 
\Gamma^{\ 0} \ ( \ R \ \rightarrow \ 2 \pi^{0} \ )
\ =
\ \frac{2}{9}
\ \alpha^{\ 3}
\ \sqrt{\ m_{\ \pi_{+}}^{2} - m_{\ \pi_{0}}^{2} \ }
\ \left | 
\ m_{\ \pi_{\ +}} 
	\ \Delta \ a_{\ D}
\ \right |^{\ 2}
\end{array}
\end{equation}
\noindent
For the purpose of getting orders of magnitude into perspective we
define the reference width, substituting the quantity $\Delta \ a$
for $\Delta \ a_{\ D}$ in eq. (\ref{eq:30})
\begin{equation}
  \label{eq:31}
      \begin{array}{l} 
      \begin{array}{lll} 
\Gamma_{\ a} \ ( \ R \ \rightarrow \ 2 \pi^{0} \ )
\ & = &
\ \frac{2}{9}
\ \alpha^{\ 3}
\ \sqrt{\ m_{\ \pi_{+}}^{2} - m_{\ \pi_{0}}^{2} \ }
\ \left | 
\ m_{\ \pi_{\ +}} 
	\ \Delta \ a
\ \right |^{\ 2}
        \vspace*{0.5cm} \\
\Gamma_{\ ref} 
\ & = &
\ \frac{1}{72}
\ \alpha^{\ 3}
\ \sqrt{\ m_{\ \pi_{+}}^{2} - m_{\ \pi_{0}}^{2} \ }
        \vspace*{0.5cm} \\
\Gamma_{\ a} \ ( \ R \ \rightarrow \ 2 \pi^{0} \ )
\ & = &
\ \Gamma_{\ ref}
\ \left ( 
\hspace*{0.2cm}
      \begin{array}{c} 
        m_{\ \pi_{\ +}} \ \Delta \ a
        \vspace*{0.2cm} \\
        \hline \vspace*{-0.2cm} \\
        0.25
      \end{array}
\hspace*{0.2cm}
\ \right )^{\ 2}
\end{array}
\end{array}
\end{equation}
\noindent
Substituting the reference width eq. (\ref{eq:30}) becomes
\begin{equation}
  \label{eq:32}
      \begin{array}{l} 
      \begin{array}{lll} 
\Gamma^{\ 0} \ ( \ R \ \rightarrow \ 2 \pi^{0} \ )
 & = &
\ \Gamma_{\ ref}
\ \left ( 
\hspace*{0.2cm}
      \begin{array}{c} 
        m_{\ \pi_{\ +}} \ \Delta \ a
        \vspace*{0.2cm} \\
        \hline \vspace*{-0.2cm} \\
        0.25
      \end{array}
\hspace*{0.2cm}
\ \right )^{\ 2}
\ \left ( 
\hspace*{0.2cm}
      \begin{array}{c} 
        \Delta \ a_{\ D}
        \vspace*{0.2cm} \\
        \hline \vspace*{-0.2cm} \\
        \Delta \ a
      \end{array}
\hspace*{0.2cm}
\ \right )^{\ 2}
        \vspace*{0.5cm} \\
\Gamma_{\ ref} 
 & = &
\ ( \ 5.397 \ ) \ 10^{\ -9} 
\ \sqrt{\ m_{\ \pi_{+}}^{2} - m_{\ \pi_{0}}^{2} \ }
\ = \ 0.1917 \ \mbox{eV}
        \vspace*{0.5cm} \\
\tau_{\ ref} 
\ = \ 1 \ / \ \Gamma_{\ ref} 
 & = &
 \ ( \ 3.434 \ ) \ 10^{\ -15} \ \mbox{sec}
\end{array}
\end{array}
\end{equation}
\noindent
We focus on the limiting behaviour of the quantities
$ \left \langle \ \varphi_{\ \vec{k}}^{\ +} \ |
\ H_{\ 1} \ | \ R \ ; \ rel \ \right \rangle $ in eq. (\ref{eq:8}) in comparison with
the scattering amplitude  
$ \left \langle \ \varphi_{\ \vec{k}}^{\ +} \ |
\ H_{\ 1} \ | \ \psi_{\ \vec{p}} \ \right \rangle $ in eq. (\ref{eq:13})
and their limiting relation in eq. (\ref{eq:18})
\begin{equation}
  \label{eq:33}
      \begin{array}{l} 
      \begin{array}{lll} 
\left \langle \ \varphi_{\ \vec{k}}^{\ +} \ |
\ H_{\ 1} \ | \ R \ ; \ rel \ \right \rangle 
\ & \rightarrow &
- \ \frac{2}{3} 
\hspace*{0.2cm}
      \begin{array}{c} 
        4 \ \pi
        \vspace*{0.2cm} \\
        \hline \vspace*{-0.2cm} \\
        m_{\ \pi_{\ +}}
      \end{array}
\hspace*{0.2cm}
	\ \chi_{\ C}
        \ \Delta \ a_{\ D} 
        \vspace*{0.5cm} \\
\left \langle \ \varphi_{\ \vec{k}}^{\ +} \ |
\ H_{\ 1} \ | \ \psi_{\ \vec{p}} \ \right \rangle \ \chi_{\ C}
\ & \rightarrow &
- \ \frac{2}{3} 
\hspace*{0.2cm}
      \begin{array}{c} 
        4 \ \pi
        \vspace*{0.2cm} \\
        \hline \vspace*{-0.2cm} \\
        m_{\ \pi_{\ +}}
      \end{array}
\hspace*{0.2cm}
	\ \chi_{\ C}
        \ \Delta \ a 
\end{array}
\end{array}
\end{equation}
\noindent
Follwoing S. Deser et al. \cite{DGBT} we consider
the purely Coulombic bound state of $\pi^{\ +} \ \pi^{\ -}$ , which
we shall denote 
$ \left | \ \psi_{\ C} \ \right \rangle $ as unperturbed state 
together with the limiting perturbing Hamiltonian
\begin{equation}
  \label{eq:34}
      \begin{array}{l} 
      H_{\ 1} \ \rightarrow \ H_{\ 1}^{\ str.}
\end{array}
\end{equation}
\noindent
and the eigenstate(s) 
$ \left | \ \Psi \ \right \rangle $
of the full Hamiltonian 
\begin{equation}
  \label{eq:35}
      \begin{array}{l} 
      H \ = \ H_{\ C} \ + \ H_{\ 1}^{\ str.}
    \hspace*{0.3cm} ; \hspace*{0.3cm}
    H_{\ C} \ = \ H_{\ 0} \ + \ V_{\ C}
\end{array}
\end{equation}
\noindent
From the two equations 
\begin{equation}
  \label{eq:36}
      \begin{array}{l} 
      H \ \left | \ \Psi \ \right \rangle
      \ = 
      \ E \ \left | \ \Psi \ \right \rangle
    \hspace*{0.3cm} ; \hspace*{0.3cm}
      H_{\ C} \ \left | \ \psi_{\ C} \ \right \rangle 
      \ = 
      \ E_{\ C} \ \left | \ \psi_{\ C} \ \right \rangle 
        \vspace*{0.5cm} \\
	\Delta \ E \ = \ E \ - \ E_{\ C}
    \hspace*{0.3cm} ; \hspace*{0.3cm}
    E_{\ C} \ = \ 2 \ m_{\ \pi_{\ +}} \ - \ \frac{1}{4} \ \alpha^{\ 2}
    \ m_{\ \pi_{\ +}}
\end{array}
\end{equation}
\noindent
the limiting relation for the energy shift $\Delta \ E$ of the resonance
follows
\begin{equation}
  \label{eq:37}
      \begin{array}{l} 
	\Delta \ E \ = 
\hspace*{0.2cm}
      \begin{array}{c} 
       \left \langle \ \psi_{\ C} \ \right |
      \ H_{\ 1}^{\ str.}
      \ \left | \ \Psi \ \right \rangle
        \vspace*{0.2cm} \\
        \hline \vspace*{-0.2cm} \\
       \left \langle \ \psi_{\ C} \ \right |
       \left . \ \Psi \ \right \rangle
      \end{array}
\end{array}
\end{equation}
\noindent
In the limit we are considering the energy shift becomes
\begin{equation}
  \label{eq:38}
      \begin{array}{l} 
      \begin{array}[t]{lll} 
	\Delta \ E \ & = &
       \ \left \langle \ \psi_{\ C} \ \right |
      \ H_{\ 1}^{\ str.}
      \ + \ P 
      \ \left (
\hspace*{0.2cm}
      \begin{array}{c} 
      1
        \vspace*{0.2cm} \\
        \hline \vspace*{-0.2cm} \\
      E_{\ C} \ - \ H 
      \end{array}
\hspace*{0.2cm}
      \ \right )
      \ H_{\ 1}^{\ str.}
      \ \left | \ \psi_{\ C} \ \right \rangle
        \vspace*{0.5cm} \\
	\ & = &
	\ - \ \chi_{\ C}^{\ 2}
\hspace*{0.2cm}
      \begin{array}{c} 
      1
        \vspace*{0.2cm} \\
        \hline \vspace*{-0.2cm} \\
      4 \ m_{\ \pi_{\ +}}^{\ 2}
      \end{array}
\hspace*{0.2cm}
\ Re \ T \ ( \ \pi^{\ +} \ \pi^{\ -} \ ; \ \pi^{\ +} \ \pi^{\ -} \ )
        \vspace*{0.5cm} \\
	\ & = &
	\ - \ \alpha^{\ 3}
\hspace*{0.2cm}
      \begin{array}{c} 
      m_{\ \pi_{\ +}}
        \vspace*{0.2cm} \\
        \hline \vspace*{-0.2cm} \\
      6 
      \end{array}
\hspace*{0.2cm}
      \ m_{\ \pi_{\ +}}
      \ ( \ 2 \ a^{\ I \ = \ 0} \ + \ a^{\ I \ = \ 2} \ )
      \end{array}
\end{array}
\end{equation}
\noindent
It is instructive to follow S. Deser et al. \cite{DGBT} and
extend the relations in eq. (\ref{eq:38})
to the (complex) elastic $\pi^{\ +} \ \pi^{\ -}$ scattering amplitude 
\begin{equation}
  \label{eq:39}
      \begin{array}{l} 
      P 
      \ \left (
\hspace*{0.2cm}
      \begin{array}{c} 
      1
        \vspace*{0.2cm} \\
        \hline \vspace*{-0.2cm} \\
      E_{\ C} \ - \ H 
      \end{array}
\hspace*{0.2cm}
      \ \right )
\ \rightarrow 
\hspace*{0.2cm}
      \begin{array}{c} 
      1
        \vspace*{0.2cm} \\
        \hline \vspace*{-0.2cm} \\
      E_{\ C} \ - \ H \ - \ i \ \varepsilon
      \end{array}
\hspace*{0.2cm}
\ \rightarrow 
        \vspace*{0.5cm} \\
	\Delta \ E \ - \ i \ \gamma \ / \ 2
	\ =
	\ - \ \chi_{\ C}^{\ 2}
\hspace*{0.2cm}
      \begin{array}{c} 
      1
        \vspace*{0.2cm} \\
        \hline \vspace*{-0.2cm} \\
      4 \ m_{\ \pi_{\ +}}^{\ 2}
      \end{array}
\hspace*{0.2cm}
\ T \ ( \ \pi^{\ +} \ \pi^{\ -} \ ; \ \pi^{\ +} \ \pi^{\ -} \ )
\ \rightarrow 
        \vspace*{0.5cm} \\
	\gamma \ = 
\ \Gamma_{\ a} \ ( \ R \ \rightarrow \ 2 \pi^{0} \ )
\end{array}
\end{equation}
\noindent
The quantity $\gamma \ = \ \Gamma_{\ a} \ ( \ R \ \rightarrow \ 2 \pi^{0} \ )$
in eqs. (\ref{eq:31}) and (\ref{eq:39}) is the limiting width of the state
$\left | \ \psi_{\ C} \ \right \rangle$ as it decays, to all orders in
the strong interaction, to $2 \ \pi^{\ 0}$ .
\vspace*{0.1cm} 

\noindent
This corresponds to the substitution in eq. (\ref{eq:33})
\begin{equation}
  \label{eq:40}
      \begin{array}{l} 
      \begin{array}{lll} 
\left | \ R \ ; \ rel \ \right \rangle 
\ & \rightarrow &
\ \left | \ \psi_{\ C} \ \right \rangle
        \vspace*{0.5cm} \\
\left \langle \ \varphi_{\ \vec{k}}^{\ +} \ |
\ H_{\ 1} \ | \ R \ ; \ rel \ \right \rangle 
\ & \rightarrow &
\left \langle \ \varphi_{\ \vec{k}}^{\ +} \ |
\ H_{\ 1} \ | \ \psi_{\ C} \ \right \rangle
        \vspace*{0.5cm} \\
\Delta \ a_{\ D} 
\ & \rightarrow &
\ \Delta \ a 
\end{array}
\end{array}
\end{equation}
\noindent
However the state $\left | \ R \ ; \ rel \ \right \rangle$ 
as it evolves from without external perturbation to a time $t \ = \ 0$
say, is not $ \left | \ \psi_{\ C} \ \right \rangle $ , but rather the full
incoming Schr\"{o}dinger state $\left | \ \varphi^{\ -} \ ; C \ \right \rangle$ ,
adiabatically evolving from $ \left | \ \psi_{\ C} \ \right \rangle $ 
\begin{equation}
  \label{eq:41}
      \begin{array}{l} 
      \left | \ \varphi^{\ -} \ ; C \ \right \rangle
      \ =
      \ \left | \ \psi_{\ C} \ \right \rangle 
      \ +
\hspace*{0.2cm}
      \begin{array}{c} 
      1
        \vspace*{0.2cm} \\
        \hline \vspace*{-0.2cm} \\
      E_{\ R} \ - \ H \ - \ i \ \varepsilon
      \end{array}
\hspace*{0.2cm}
      \ H_{\ 1}^{\ str.}
      \ \left | \ \psi_{\ C} \ \right \rangle 
\end{array}
\end{equation}
\noindent
with the energy $E_{\ R}$ including the energy shift $\Delta \ E$ 
in eq. (\ref{eq:38}) . 
\vspace*{0.1cm} 

\noindent
In the limit we are considering it follows
\begin{equation}
  \label{eq:42}
      \begin{array}{l} 
      \begin{array}{lll} 
      \left | \ \varphi^{\ -} \ ; \ C \ \right \rangle
 & \rightarrow &
 \chi_{\ C} 
      \left | \ \varphi^{\ -} \ ; \ \pi^{\ +} \ \pi^{\ -} \ \right \rangle
        \vspace*{0.5cm} \\
\left \langle \ \varphi^{\ +} \ |
\ H_{\ 1} \ | \ R \ ; \ rel \ \right \rangle 
 & \rightarrow &
      \begin{array}[t]{l} 
\ \chi_{\ C} 
        \vspace*{0.5cm} \\
\ \left \langle \ \varphi^{\ +} \ ; \ 2 \ \pi^{\ 0} \ |
\ H_{\ 1} \ |  
       \ \varphi^{\ -} \ ; \ \pi^{\ +} \ \pi^{\ -} \ \right \rangle
\end{array}
\end{array}
        \vspace*{0.5cm} \\
\vec{k} \ = \ \vec{p} \ = \ 0
    \hspace*{0.3cm} ; \hspace*{0.3cm}
    H_{\ 1}^{\ str.} \ \rightarrow \ H_{\ 1}
\end{array}
\end{equation}
\noindent
From the structure of limiting amplitudes in eq. (\ref{eq:42})
we infer , dropping the superscript $^{str.}$ in the following 
\begin{equation}
  \label{eq:43}
      \begin{array}{l} 
\ \left \langle \ \varphi^{\ +} \ ; \ " 2 \ \pi^{\ 0} "\ |
\ H_{\ 1} \ |  
       \ \varphi^{\ -} \ ; \ \pi^{\ +} \ \pi^{\ -} \ \right \rangle
       \ = 
        \vspace*{0.5cm} \\
\hspace*{1.0cm} \left \langle \ \psi \ ; \ " 2 \ \pi^{\ 0} "\ |
\ H_{\ 1} 
\ \left (
\hspace*{0.2cm}
      \begin{array}{c} 
      1
        \vspace*{0.2cm} \\
        \hline \vspace*{-0.2cm} \\
      1 \ - \ G_{\ 0} \ H_{\ 1}
      \end{array}
\hspace*{0.2cm}
\ \right )^{\ 2}
\ | \ \psi \ ; \ \pi^{\ +} \ \pi^{\ -} \ \right \rangle
        \vspace*{0.5cm} \\
G_{\ 0} \ = 
\ \left ( \ E_{\ thr} \ - \ H_{\ 0} \ + \ i \ \varepsilon \ \right )^{\ -1}
\end{array}
\end{equation}
\noindent
We recall that $\psi$ in eq. (\ref{eq:43}) refers to plane wave states
with vanishing momentum at threshold.
\vspace*{0.1cm} 

\noindent
The argument for the limiting substitutions in eq. (\ref{eq:42}) is
like this : in the matrix element 
$\left \langle \ \varphi^{\ +} \ | \ H_{\ 1} \ | \ R \ ; \ rel \ \right \rangle$
the action of $H_{\ 1} \ \rightarrow \ H_{\ 1}^{\ str.}$ restricts the
configurations composing both states  
$\left | \ \varphi^{\ +} \ \ \right \rangle$ and
$\left | \ R \ ; \ rel \ \right \rangle$ to normal strong interaction relative
distances $d_{\ rel} \ \leq \ \sim \ 1 \ - \ 4 \ \mbox{fm}$. 

\noindent
Those configurations are in the sense of the limit considered insensitive
to the two key parameters governing the resonance at its determining
distance, i.e. its Bohr radius 

\noindent
$a_{\ B} \ = \ 2 \ / \ ( \ m_{\ \pi_{\ +}} \ \alpha \ ) \ \sim \ 400 \ \mbox{fm}$
\vspace*{0.1cm} 

\noindent
As a consequence also the key mass square difference

\noindent
$m_{\ \pi_{\ +}}^{\ 2} - \ m_{\ \pi_{\ 0}}^{\ 2} \ \sim 
\ ( \ 35.51 \ \mbox{MeV} \ )^{\ 2}$ 

\noindent
plays no significant role.
Hence the dominant configurations relevant for the transition amplitude
within the state $\left | \ R \ ; \ rel \ \right \rangle$ are the same as those
in which R is (almost) bound , i.e. for $\alpha \ \neq \ 0$ but
$m_{\ \pi_{\ +}} \ = \ m_{\ \pi_{\ 0}}$ . Of course the resonance R is always
decaying into two (and more) photons.
\vspace*{0.1cm} 

\noindent
In the above situation the (almost) bound state R is by no means described by the
Coulomb wave function, in particular at the distances within $d_{\ rel}$ .
\vspace*{0.1cm} 

\noindent
It follows combining eqs. (\ref{eq:43}) and (\ref{eq:33})
\begin{equation}
  \label{eq:44}
      \begin{array}{l} 
- \ \Delta \ a_{\ D} 
\ \rightarrow 
        \vspace*{0.5cm} \\
 \hspace*{-0.2cm}
      \begin{array}{c} 
        3 \ m_{\ \pi_{\ +}}
        \vspace*{0.2cm} \\
        \hline \vspace*{-0.2cm} \\
        8 \ \pi
      \end{array}
\hspace*{0.2cm}
\left \langle \ \psi \ ; \ 2 \ \pi^{\ 0} \ |
\ H_{\ 1} 
\ \left (
\hspace*{0.2cm}
      \begin{array}{c} 
      1
        \vspace*{0.2cm} \\
        \hline \vspace*{-0.2cm} \\
      1 \ - \ G_{\ 0} \ H_{\ 1}
      \end{array}
\hspace*{0.2cm}
\ \right )^{\ 2}
\ | \ \psi \ ; \ \pi^{\ +} \ \pi^{\ -} \ \right \rangle
        \vspace*{0.5cm} \\
- \ \Delta \ a 
\ \rightarrow 
        \vspace*{0.5cm} \\
 \hspace*{-0.2cm}
      \begin{array}{c} 
        3 \ m_{\ \pi_{\ +}}
        \vspace*{0.2cm} \\
        \hline \vspace*{-0.2cm} \\
        8 \ \pi
      \end{array}
\hspace*{0.2cm}
\left \langle \ \psi \ ; \  2 \ \pi^{\ 0} \ |
\ H_{\ 1} 
\ \left (
\hspace*{0.2cm}
      \begin{array}{c} 
      1
        \vspace*{0.2cm} \\
        \hline \vspace*{-0.2cm} \\
      1 \ - \ G_{\ 0} \ H_{\ 1}
      \end{array}
\hspace*{0.2cm}
\ \right )
\hspace*{0.2cm}
\ | \ \psi \ ; \ \pi^{\ +} \ \pi^{\ -} \ \right \rangle
\end{array}
\end{equation}
\noindent
Eq. (\ref{eq:44}) shows that decay and scattering amplitudes are not the same.
\vspace*{0.1cm} 

\noindent
$\Delta \ a_{\ D}$ and $\Delta \ a$ according to eq. (\ref{eq:44}) can not
be related to each other without detailed knowledge of $H_{\ 1}$ .

\noindent
Let us introduce the coupling strength $\lambda$ - always remaining in
the 2 flavor SU2 symmetric QCD limit, with fixed 
$m_{\ \pi_{\ +}} \ = \ m_{\ \pi_{\ 0}}$ - through the substitution 
\begin{equation}
  \label{eq:45}
      \begin{array}{l} 
      \begin{array}{lll} 
      H_{\ 0} \ , \ G_{\ 0}
\ & \rightarrow &
      \ H_{\ 0} \ , \ G_{\ 0}
        \vspace*{0.5cm} \\
	H_{\ 1}
\ & \rightarrow &
	\ \lambda \ H_{\ 1}
        \vspace*{0.5cm} \\
\Delta \ a_{\ D} \ , \ \Delta \ a
\ & \rightarrow &
\ \Delta \ a_{\ D} \ ( \ \lambda \ ) \ , \ \Delta \ a \ ( \ \lambda \ )
\end{array}
\end{array}
\end{equation}
\noindent
Then it follows from eq. (\ref{eq:44})
\begin{equation}
  \label{eq:46}
      \begin{array}{l} 
\Delta \ a_{\ D} \ ( \ \lambda \ ) \ = 
\ \lambda 
\hspace*{0.2cm}
      \begin{array}{c} 
      d
        \vspace*{0.2cm} \\
        \hline \vspace*{-0.2cm} \\
      d \ \lambda
      \end{array}
\hspace*{0.2cm}
\ \Delta \ a \ ( \ \lambda \ )
\end{array}
\end{equation}
\noindent
In the limit considered the quantities 
$\Delta \ a_{\ D} \ , \ \Delta \ a$ depend within QCD on the two
basic parameters $\Lambda_{\ 2 \ QCD}$ and $m_{\ u} \ = \ m_{\ d}$ .
\vspace*{0.1cm} 

\noindent
An equivalent set is $f_{\ \pi}$ , the pion decay constant , and
$m_{\ \pi_{\ +}} \ = \ m_{\ \pi_{\ 0}}$. It follows from the relation in eq.
(\ref{eq:46}) that the pion mass is to be held constant, whereas at least
in lowest two orders of chiral perturbation theory \cite{Gasser} 
the variation of $\lambda$ is equivalent to a variation of $f_{\ \pi}^{\ -2}$
\begin{equation}
  \label{eq:47}
      \begin{array}{l} 
 \lambda 
\hspace*{0.2cm}
      \begin{array}{c} 
      d
        \vspace*{0.2cm} \\
        \hline \vspace*{-0.2cm} \\
      d \ \lambda
      \end{array}
\hspace*{0.2cm}
\ \sim 
\hspace*{0.2cm}
\ \varrho 
\hspace*{0.2cm}
      \begin{array}{c} 
      d
        \vspace*{0.2cm} \\
        \hline \vspace*{-0.2cm} \\
      d \ \varrho
      \end{array}
\hspace*{0.2cm}
        \vspace*{0.5cm} \\
	\varrho \ = \ k 
\hspace*{0.2cm}
      \begin{array}{c} 
      m_{\ \pi}^{\ 2}
        \vspace*{0.2cm} \\
        \hline \vspace*{-0.2cm} \\
      f_{\ \pi}^{\ 2}
      \end{array}
\hspace*{0.2cm}
    \hspace*{0.3cm} ; \hspace*{0.3cm}
    k \ \mbox{arbitrary fixed constant}
\end{array}
\end{equation}
\vspace*{0.1cm} 

\noindent
{\bf Estimates of pionium lifetime}
\vspace*{0.1cm} 

\noindent
We use eqs. (\ref{eq:46}) and (\ref{eq:47}) to {\em estimate}
$\Delta \ a_{\ D}$ and the lifetime of pionium, with the
shorthand notation $m_{\ \pi_{\ +}} \ \rightarrow \ m_{\ \pi}$

\noindent
The lowest order (tree level) values \cite{Weinb} are
\begin{equation}
  \label{eq:48}
      \begin{array}{l} 
      m_{\ \pi} \ a^{\ I \ = \ 0}_{\ (1)} \ =
\hspace*{0.4cm}
      \begin{array}{c} 
      7 \ m_{\ \pi}^{\ 2}
        \vspace*{0.2cm} \\
        \hline \vspace*{-0.2cm} \\
      32 \ \pi \ f_{\ \pi}^{\ 2}
      \end{array}
\hspace*{0.2cm}
\ = \ 0.1562
        \vspace*{0.5cm} \\
      m_{\ \pi} \ a^{\ I \ = \ 2}_{\ (1)} \ =
      \ - 
\hspace*{0.2cm}
      \begin{array}{c} 
      2 \ m_{\ \pi}^{\ 2}
        \vspace*{0.2cm} \\
        \hline \vspace*{-0.2cm} \\
      32 \ \pi \ f_{\ \pi}^{\ 2}
      \end{array}
\hspace*{0.2cm}
\ = \ - 0.0446
\end{array}
\end{equation}
\vspace*{0.1cm} 

\noindent
To this end we list from Ecker et al. \cite{EBG}
the contributions through two loop order to both scattering
lengths $a^{\ I \ = \ 0}$ and $a^{\ I \ = \ 2}$ 
\begin{equation}
  \label{eq:49}
      \begin{array}{l} 
      \begin{array}{llll} 
\mbox{order}   & m_{\ \pi} \ a^{\ I \ = \ 0} &
      m_{\ \pi} \ a^{\ I \ = \ 2} & m_{\ \pi} \ \Delta \ a
      \vspace*{0.2cm} \\ 
      \hline \\
      1                   & 0.16  & - 0.045  & 0.205
      \vspace*{0.3cm} \\
      2                   & 0.04  & \hspace*{0.3cm} 0.003  & 0.037
      \vspace*{0.3cm} \\
      3 \ (I)             & 0.017 & \hspace*{0.3cm} 0.0007 & 0.0163
      \vspace*{0.3cm} \\
      3 \ (II)            & 0.006 &  - 0.0023 & 0.0083
      \vspace*{0.3cm} \\
      \mbox{total} \ (I)  & 0.217 & - 0.0413 & 0.258
      \vspace*{0.3cm} \\
      \mbox{total} \ (II) & 0.206 & - 0.0443 & 0.250
\end{array}
\end{array}
\end{equation}
\noindent
We base our estimate on the one loop contribution to 
$a^{\ I \ = \ 0}$ which is of the form
\begin{equation}
  \label{eq:50}
      \begin{array}{l} 
a^{\ I \ = \ 0}_{\ (2)} \ = \ k_{\ 1} \ \varrho^{\ 2} \ L 
    \hspace*{0.3cm} ; \hspace*{0.3cm}
    L \ = \ - \ \log \ \varrho \ \sim \ 4
\end{array}
\end{equation}
The numerical value of the logarithm in the one loop contribution to
$a^{\ I \ = \ 0}$ is quite accurately 4 as in eq. (\ref{eq:50}) , when
all nonlogarithmic terms of order $ \varrho^{\ 2}$ are absorbed into the
argument of the logarithm. The constant k in eq. (\ref{eq:47}) can then
be chosen such that $\varrho \ = \exp \ -4$ according to eq. (\ref{eq:50}) .
\vspace*{0.1cm} 

\noindent
It then follows
\begin{equation}
  \label{eq:51}
      \begin{array}{l} 
 \lambda 
\hspace*{0.2cm}
      \begin{array}{c} 
      d
        \vspace*{0.2cm} \\
        \hline \vspace*{-0.2cm} \\
      d \ \lambda
      \end{array}
\hspace*{0.2cm}
a^{\ I \ = \ 0}_{\ (2)} \ = 
\ \left ( \ 1 \ - \ L^{\ -1} \ \right ) 
\ a^{\ I \ = \ 0}_{\ (2)} 
\ \sim \ \frac{3}{4}
\ a^{\ I \ = \ 0}_{\ (2)} \ = \ 0.03
\end{array}
\end{equation}
\noindent
Neglecting all other corrections from higher orders and from the
I = 2 channel we obtain as our estimate
\begin{equation}
  \label{eq:52}
      \begin{array}{l} 
      \Delta \ a_{\ D} \ \sim \ 0.28
    \hspace*{0.3cm} ; \hspace*{0.3cm}
      \Delta \ a \ \sim \ 0.25
    \hspace*{0.3cm} \rightarrow \hspace*{0.3cm}
\hspace*{0.2cm}
      \begin{array}{c} 
      \Delta \ a_{\ D}
        \vspace*{0.2cm} \\
        \hline \vspace*{-0.2cm} \\
      \Delta \ a
      \end{array}
\hspace*{0.2cm}
\ \sim \ 1.12
\end{array}
\end{equation}
\noindent
With the ratio $\Delta \ a_{\ D} \ / \ \Delta \ a$ given in eq. (\ref{eq:52})
we obtain from eq. (\ref{eq:32})
\begin{equation}
  \label{eq:53}
      \begin{array}{l} 
      \begin{array}{lll} 
\Gamma^{\ 0} \ ( \ R \ \rightarrow \ 2 \pi^{0} \ )
 & = &
 \ 1.25
\ \Gamma_{\ ref}
\ \left ( 
\hspace*{0.2cm}
      \begin{array}{c} 
        m_{\ \pi_{\ +}} \ \Delta \ a
        \vspace*{0.2cm} \\
        \hline \vspace*{-0.2cm} \\
        0.25
      \end{array}
\hspace*{0.2cm}
\ \right )^{\ 2}
        \vspace*{0.5cm} \\
& = &
\ 0.240 \ \mbox{eV}
\ \left ( 
\hspace*{0.2cm}
      \begin{array}{c} 
        m_{\ \pi_{\ +}} \ \Delta \ a
        \vspace*{0.2cm} \\
        \hline \vspace*{-0.2cm} \\
        0.25
      \end{array}
\hspace*{0.2cm}
\ \right )^{\ 2}
\end{array}
\end{array}
\end{equation}
\noindent
or equivalently for the lifetime, using the abbreviation
$\Gamma^{\ 0} \ ( \ R \ \rightarrow \ 2 \pi^{0} \ ) \ \rightarrow \ \Gamma^{\ 0}$
\begin{equation}
  \label{eq:54}
      \begin{array}{l} 
      \begin{array}{lll} 
\tau_{\ 0} 
\ = \ 1 \ / \ \Gamma^{\ 0} 
 & = &
 \ ( \ 2.74 \ ) \ 10^{\ -15} \ \mbox{sec}
\ \left ( 
\hspace*{0.2cm}
      \begin{array}{c} 
        0.25
        \vspace*{0.2cm} \\
        \hline \vspace*{-0.2cm} \\
        m_{\ \pi_{\ +}} \ \Delta \ a
      \end{array}
\hspace*{0.2cm}
\ \right )^{\ 2}
\end{array}
\end{array}
\end{equation}
\noindent
Finally we apply the radiative corrections as estimated by 
A.Rusetsky and H.Sazdjan \cite{AR} , \cite{HSaz} 
according to eqs. (\ref{eq:25}) and (\ref{eq:26}) and find for the
resonance width the estimate
\begin{equation}
  \label{eq:55}
      \begin{array}{l} 
      \begin{array}{lll} 
\Gamma \ ( \ R \ \rightarrow \ 2 \pi^{0} \ )
 & \sim &
 \ 1.33
\ \Gamma_{\ ref}
\ \left ( 
\hspace*{0.2cm}
      \begin{array}{c} 
        m_{\ \pi_{\ +}} \ \Delta \ a
        \vspace*{0.2cm} \\
        \hline \vspace*{-0.2cm} \\
        0.25
      \end{array}
\hspace*{0.2cm}
\ \right )^{\ 2}
\hspace*{0.2cm}
      \begin{array}{c} 
        1 \ + \ \delta
        \vspace*{0.2cm} \\
        \hline \vspace*{-0.2cm} \\
        1.06
      \end{array}
\hspace*{0.2cm}
        \vspace*{0.5cm} \\
& = &
\ 0.255 \ \mbox{eV}
\hspace*{0.3cm} \left ( 
\hspace*{0.2cm}
      \begin{array}{c} 
        m_{\ \pi_{\ +}} \ \Delta \ a
        \vspace*{0.2cm} \\
        \hline \vspace*{-0.2cm} \\
        0.25
      \end{array}
\hspace*{0.2cm}
\ \right )^{\ 2}
\hspace*{0.2cm}
      \begin{array}{c} 
        1 \ + \ \delta
        \vspace*{0.2cm} \\
        \hline \vspace*{-0.2cm} \\
        1.06
      \end{array}
\hspace*{0.2cm}
\end{array}
\end{array}
\end{equation}
\noindent
or equivalently for the lifetime
\begin{equation}
  \label{eq:56}
      \begin{array}{l} 
\tau \ ( \ R \ \rightarrow \ 2 \pi^{0} \ )
\ = 
 \ ( \ 2.58 \ ) \ 10^{\ -15} \ \mbox{sec}
\ \left ( 
\hspace*{0.2cm}
      \begin{array}{c} 
        0.25
        \vspace*{0.2cm} \\
        \hline \vspace*{-0.2cm} \\
        m_{\ \pi_{\ +}} \ \Delta \ a
      \end{array}
\hspace*{0.2cm}
\ \right )^{\ 2}
\hspace*{0.2cm}
      \begin{array}{c} 
        1.06
        \vspace*{0.2cm} \\
        \hline \vspace*{-0.2cm} \\
        1 \ + \ \delta
      \end{array}
\hspace*{0.2cm}
\end{array}
\end{equation}
\noindent
If we omit the correction proportional to 
$( \ \Delta \ a_{\ D} \ / \ \Delta \ a \ )^{\ 2}$
the corresponding estimate for the lifetime would be
\begin{equation}
  \label{eq:57}
      \begin{array}{l} 
\tau \ ( \ R \ \rightarrow \ 2 \pi^{0} \ )
\ = 
 \ ( \ 3.24 \ ) \ 10^{\ -15} \ \mbox{sec}
\ \left ( 
\hspace*{0.2cm}
      \begin{array}{c} 
        0.25
        \vspace*{0.2cm} \\
        \hline \vspace*{-0.2cm} \\
        m_{\ \pi_{\ +}} \ \Delta \ a
      \end{array}
\hspace*{0.2cm}
\ \right )^{\ 2}
\hspace*{0.2cm}
      \begin{array}{c} 
        1.06
        \vspace*{0.2cm} \\
        \hline \vspace*{-0.2cm} \\
        1 \ + \ \delta
      \end{array}
\hspace*{0.2cm}
\end{array}
\end{equation}
\vspace*{0.1cm} 

\noindent
The results on the above lifetime to date are from L. Nemenov et al. \cite{Neme}
\begin{equation}
  \label{eq:58}
      \begin{array}{l} 
\tau \ ( \ R \ \rightarrow \ 2 \pi^{0} \ )
\ = 
 \ ( \ 2.9 \ ^{+ \ \infty}_{- \ 2.1} \ ) \ 10^{\ -15} \ \mbox{sec}
\end{array}
\end{equation}
\vspace*{0.1cm} 

\noindent
In conclusion we are looking forward in suspense to the measurement or better the
analytic deduction of the lifiteme of pionium from
the study of the breakup reaction in targets with appropriately chosen thickness
by the DIRAC collaboration \cite{DIRAC} .


\vspace*{0.3cm} 

\newpage

\noindent
\large{\bf Acknowledgment}
\vspace*{0.1cm} 
 
\noindent
I should like to thank the organizers of this workshop for their large effort, which 
created a unique atmosphere of intense discussion and learning.











\end{document}